# Highly Nonlinear Luminescence Induced by Gold Nanoparticles on Glass Surfaces with Continuous-Wave Laser Illumination

Yong Wu[*], Xundong Wu, Ligia Toro, and Enrico Stefani

[*]BH-428 CHS, University of California, Los Angeles, CA, 90095-7115. ywu.thu@gmail.com

**Key words**: nanoparticles, nonlinear optics, multiphoton microscopy, super-resolution

## Abstract

We report on highly nonlinear luminescence being observed from individual spherical gold nanoparticles immobilized on a glass surface and illuminated by continuous-wave (CW) lasers with relatively low power. The nonlinear luminescence shows optical super-resolution beyond the diffraction limit in three dimensions compared to the scatting of the excitation laser light. The luminescence intensity from most nanoparticles is proportional to the $5^{th}$—$7^{th}$ power of the excitation laser power and has wide excitation and emission spectra across the visible wavelength range. Strong nonlinear luminescence is 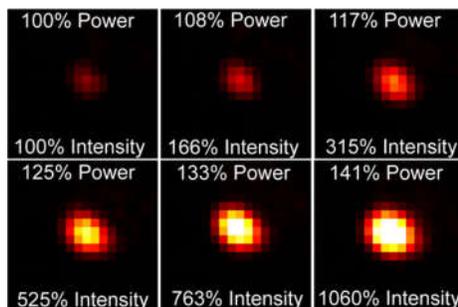 only observed near the glass surface. High optical nonlinearity excited by low CW laser power is related to a long-lived dark state of the gold nanoparticles, where the excitation light is strongly absorbed. This phenomenon has potential biological applications in super-resolution and deep tissue imaging.

## Introduction

Optical nonlinearity induced by metal nanomaterials has been intensively studied. For perfectly spherical nanoparticles, the lowest-order nonlinear effect is the third-harmonic generation, which was observed for individual gold nanoparticles using a picosecond pulsed laser source [1]. Metal (especially gold and silver) nanoparticles or nanoclusters can also be made luminescent or fluorescent [2-5] and used to for biomedical applications such as live cell imaging for its low toxicity [6] and correlative microscopy combining photothermal and fluorescence techniques [7]. Luminescence from metal nanoparticles can be induced by multiphoton absorption [8], which has potential applications in deep tissue imaging . Strong electromagnetic field induced by plasmon resonance of metal nanoparticles/nanoclusters/nanorods has been widely used to enhance the signal (e.g., fluorescence, Raman scattering or four-wave mixing) level [9-11] or to dramatically increase the probability of multiphoton processes in the environment [12].

In this paper, we report our observation of highly nonlinear luminescence emission from gold nanoparticles immobilized on borosilicate glass surfaces. Compared with previous reported phenomena, our observation has the following features: (1) the nonlinearity is unusually strong. The luminescence intensity is proportional to the $5^{th}$—$7^{th}$ power of the excitation laser irradiance, whereas in most previously studies the nonlinearity was up to the $3^{rd}$ order; (2) such strong nonlinearity can be induced by less than 5 mW of continuous-wave (CW) excitation power for gold nanoparticles with a 70 nm diameter, while previous studies usually use fast pulsed laser sources to attain high peak irradiance; (3) the nonlinear luminescence we have observed can occur far away from the plasmon resonance peak of the gold nanoparticles.

**Materials and Methods**

Colloidal gold nanoparticles with diameter ranging from 1.8 nm to 70 nm were purchased from Nanopartz Inc., CO, USA. Experiments results presented in this paper were acquired using 70 nm nanoparticles because the nonlinear luminescence can be observed with lower excitation laser power, but similar phenomena were observed for smaller particles as well. Gold nanoparticles we used were either bare or conjugated to IgG antibodies, with concentration varying from 0.05 mg/ml to 5 mg/ml. We did not notice any differences in results due to the above variations in our experiments. Glass coverslips (CS-12R17, Warner Instruments, CT, USA) were made from borosilicate glass that is commonly used in fluorescence microscopy.

To make a gold nanoparticle sample, colloidal gold nanoparticles were well shaken and sonicated for at least half an hour. A drop of gold colloid was then placed between two pieces of glass coverslips forming a liquid membrane. After drying in air for 2—3 hours, water evaporated and the gold nanoparticles were immobilized on the glass surface. The two coverslips were then stuck together by a drop of mounting medium (Prolong Gold, Life Technologies, NY, USA). We adopted this sample preparation protocol in order to attain uniformly localized nanoparticles and to reduce excessive laser light reflection from the glass surface. However, this protocol is not essential to observe the nonlinear luminescence. For example, the nonlinear luminescence could also be seen with a spin-coated sample without using mounting media.

Luminescence and Raleigh/Mie scattering were observed using a custom-built resonant-scanning confocal microscope. The microscope has four CW diode laser lines: 405 nm (Coherent, CA, USA), 473 nm (CNI, Changchun, China), 561 nm (Melles Griot, NY, USA) and 646 nm (Crystal, NV, USA). The horizontal scanner is a resonant scanner (CRS 8 KHz, Cambridge Technology, MA, USA) and the vertical scanner is a galvanometer mirror (M2S, Cambridge Technology, MA, USA). A high-NA oil-immersion objective (Plan Apo 60X NA=1.42, Olympus, Japan) was used for its high optical resolution. Barrier filters (Semrock, NY, USA) were used to control the detection wavelength range. Since the luminescence is highly nonlinear, a pinhole is in principle unnecessary (this is similar to two photon microscopy). In practice, we used pinholes with diameter much larger than one Airy unit.

The luminescence signal can be quite intensive and the microscope acquisition system must have excellent linearity to accurately quantify the relationship between the luminescence intensity and the excitation laser power. Our acquisition system is a combination of a photomultiplier (PMT; H7422-40, Hamamatsu Photonics, Japan), a current-to-voltage amplifier (HC-130, Hamamatsu Photonics, Japan), and an image grabber (Odyssey XPro, Matrox Imaging, Quebec, Canada). The power of the laser beams was measured by a power meter (PM130D, Thorlabs, NJ, USA) at the back aperture of the objective.

**Results**

The nonlinear luminescence has a striking visual feature: it shows unusually higher optical resolution in both the lateral direction and the axial direction. This is illustrated in Fig. 1, which compares the optical resolution of the nonlinear fluorescence to the scattering of the excitation laser light. Gold nanoparticles with a diameter of 70 nm on glass were illuminated by 405 nm laser light. Fig. 1A is the image of light-scattering, taken with low laser power (~0.2 mW) and without using barrier filters. Fig. 1B was taken for the same field of view as in Fig. 1A with a much higher excitation power (3 mW; corresponding to ~3 MW/cm$^2$ in-focus irradiance) and a barrier filter (FF01-665/150, Semrock, NY, USA; the transmission window is centered at 665 nm and its width is 150 nm) to block the unwanted light-scattering and fluorescence from glass. Throughout this paper, we show light-scattering images in gray-scale and nonlinear luminescence images with the "red-hot" pseudo-color. To see the difference in resolution more clearly, a region was blown up and shown in Fig. 1C and Fig. 1D. One can see that a single light-scattering spot in Fig. 1C were resolved into 3 particles in Fig. 1D. The point spread function (PSF) of light-scattering is diffraction-limited: an image intensity line profile of light-scattering (the blue data points and the blue curve in Fig. 1E) from a gold nanoparticle (indicated by the white arrowhead in Fig. 1A) shows a full-width-half-maximum

(FWHM) of 185 nm (the theoretical resolution limit for a confocal microscope with a large pinhole for an infinitesimal object is $0.51 \cdot \lambda/NA \approx 145$ nm). In contrast, the FWHM of the PSF of the nonlinear luminescence from the same particle (the red data points and the red curve in Fig. 1E) is only 78 nm and apparently beyond the diffraction limit. Similarly, Fig. 1F—1H show that the axial resolution of nonlinear luminescence of another nanoparticle is 208 nm, whereas its PSF FWHM of light-scattering was fitted to be 577 nm (the theoretical limit is $0.88 \cdot \lambda/(n - \sqrt{n^2 - NA^2}) \approx 361$ nm).

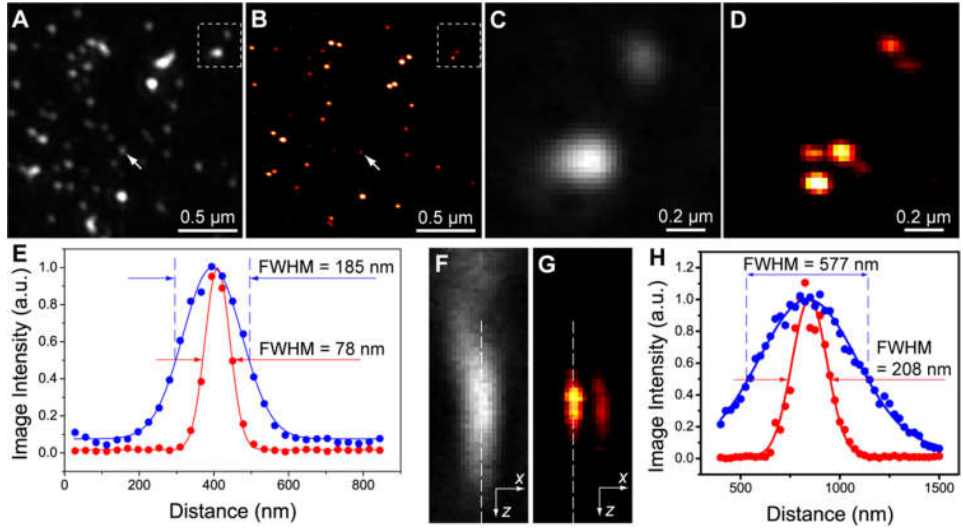

Fig. 1 Nonlinear luminescence shows super-resolution. (A) Scattering of 405 nm laser light from gold nanoparticles on glass. (B) In the same region as in A, nonlinear luminescence was excited by the same 405 nm laser with much higher optical power. (C and D) Blowups of the white squared regions in A and B, respectively. (E) Lateral resolution is quantified by fitting the intensity line-profiles for an individual particle (pointed by the white arrowheads in A and B) to Gaussian functions. PSF of luminescence (red) has a FWHM of 78 nm, compared to the diffraction-limited light-scattering PSF (blue). (F and G) Side views (x-z plane) of the light-scattering and the nonlinear luminescence focal spots of two gold nanoparticles, respectively. Only with luminescence the two particles are differentiated. (H) Axial resolution of nonlinear luminescence is 208 nm and beyond the diffraction limit.

We explain the apparent diffraction-unlimited optical resolution of the luminescence by its strong nonlinearity. For simplicity we neglect the impact of the confocal pinhole and the light collection by the objective and assume that the optical resolution is solely determined by the PSF of the excitation light. This is valid when the pinhole is larger. In a linear emission process, such as light-scattering, the optical resolution is linearly proportional to the excitation PSF. However, if the probability for the emission process to happen obeys a power law and is proportional to the $n^{th}$ power of excitation intensity, the optical resolution is then proportional to the $n^{th}$ power of the excitation PSF. The PSF can be well approximated by a Gaussian function and thus the resolution of luminescence $d_{lumin}$ is enhanced by a factor of $\sqrt{n}$ compared with the resolution of light-scattering $d_{scat}$.

$$d_{lumin} = \sqrt{n} \cdot d_{scat} \qquad (1)$$

Therefore, in the above example, the lateral resolution enhancement from 185 nm to 78 nm implies that the luminescence obeys a power law of the 5$^{th}$—6$^{th}$ order; the axial resolution enhancement from 577 nm to 208 nm implies an even higher nonlinearity to the 7$^{th}$—8$^{th}$ order. To determine this order of nonlinearity with high accuracy, we directly measured the relationship between the luminescence intensity and the excitation laser power, and the results are shown in Fig. 2. Fig. 2A is a visual display of the strong nonlinearity: by increasing the excitation power by merely 41%, the luminescence emission from a gold nanoparticle was enhanced by 9.6 times. This strong nonlinearity is also visualized by in a log-log plot shown in Fig. 2B, and the slope was fitted to 7.01±0.15, and the adjust coefficient of determination ($R^2_{adj.}$), which indicates the soundness of the fitting, is 0.9976. As a comparison, we show the relationship between light –scattering and the excitation power in Fig. 2C, in which the acquisition system was maintained to operate in the same condition as in the luminescence measurement, and the laser power was chosen such that the signal level is similar to the luminescence. The laser wavelength was chosen to be 646 nm, which is in the wavelength range where the luminescence intensity is relatively strong (see the description of the luminescence emission spectrum below). When the illumination power is very low (< 5 µW), the light-scattering is a perfectly linear process: the linear fitting of the emission intensity as a function of illumination power has an adjust coefficient of determination $R^2_{adj.} = 0.9999$. It proves that the strong nonlinearity we observed in Fig. 2B is real, not due to any artifacts in the acquisition system.

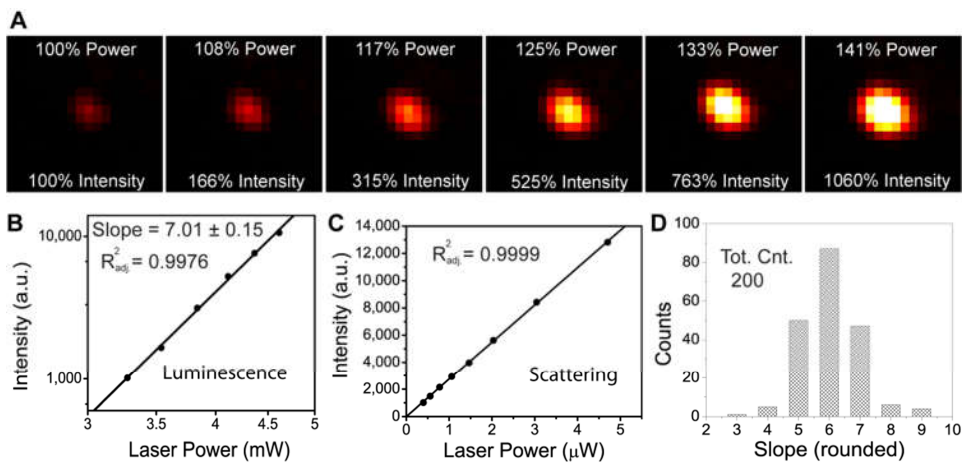

Fig. 2 Luminescence intensity is a highly nonlinear function of the excitation laser power. For example, (A) luminescence intensity of a gold nanoparticle grew 9.6 times as a response to a 41% increase in excitation power. (B) A log-log plot shows that luminescence is proportional to the 7$^{th}$ power of excitation power. (C) Light-scattering is perfectly linear, measured by the same acquisition system at a similar signal level. (D) Statistic of 200 nanoparticles shows that luminescence intensity is proportional to the 3$^{rd}$—9$^{th}$ power of excitation power. Most particles are in the range of the 5$^{th}$—7$^{th}$ power.

Note that though almost all particles show strong nonlinearity in luminescence, they do not all have such a nice fitting as shown in Fig. 2B. This is due to photobleaching and strong photoblinking of the luminescence. A real-time video of the luminescence photoblinking can be found at https://sites.google.com/site/ywurfimor/home/photo-blinking-of-nonlinear-luminescence. Also, not all particles produce luminescence with the same order of nonlinearity. We measured 483 particles, among which only 200 have excellent linear fitting for luminescence intensity as a function of the excitation power in a log-log plot (by "excellent linear fitting" we mean $R^2_{adj.} \geq 0.99$). The 200 particles display a broad distribution of the strength of the luminescence nonlinearity, measured by the fitted slope rounded to the nearest whole number, as shown in Fig. 2D. Particles are mostly likely to have an order of nonlinearity at 6, and 97% of the particles have an order no lower than 5.

The nonlinear luminescence has a wide excitation spectrum. It can be effectively excited by 405 nm, 473 nm, and 561 nm laser light, as illustrated by Fig. 3, which shows luminescence images of the same gold nanoparticles excited by the 405 nm, 473 nm and 561 nm lasers, respectively, all under the same excitation power (1.5 mW). The excitation efficiency at 405 nm and 473 nm seems to be a little higher than at 561 nm, despite that the latter is closer to the plasmon resonance peak of the nanoparticles (~550 nm according to the vendor). Also, the luminescence has a wide emission spectrum, too. This is illustrated by Fig. 4, where the luminescence was recorded using four different barrier filters, whose center wavelengths were 445 nm, 520 nm, 607 nm, and 685 nm, respectively. It is obvious that the wavelength of luminescence spans almost the entire visible band, while luminescence is most strong in the wavelength range of ~600—700.

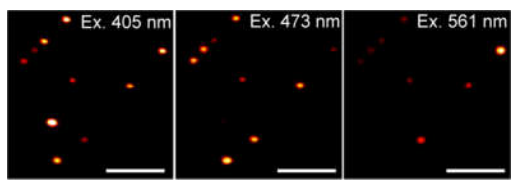

Fig. 4 Nonlinear luminescence is efficiently excited away from plasmon resonance. From left to right: 70 nm gold nanoparticles (peak plasmin resonance wavelength ~550 nm) on glass were excited by a 405 nm, 473 nm, and 561 nm laser, respectively. Images were recorded with a 665/150 (center transmission wavelength 6655 nm, FWHM 150 nm) barrier filter. Images are shown with the same contrast. Scale bars represent 1 μm.

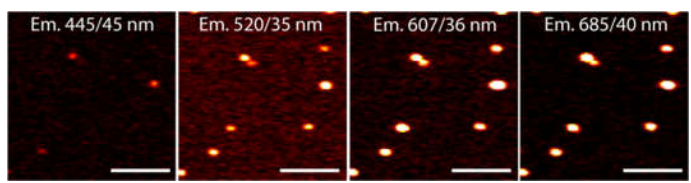

Fig. 4 Nonlinear luminescence has a wide emission spectrum. 70 nm gold nanoparticles on glass were excited by a 405 nm laser and 4 images were taken with 4 different barrier filters for the same field of view. From left to right are: 445/45 nm (center transmission wavelength 445 nm, FWHM 45 nm), 520/35 nm, 607/36 nm, and 685/40 nm. Images are shown with the same contrast. Scale bars represent 1 μm.

We only observed strong nonlinear luminescence near the glass surface, indicating that this phenomenon is closely related to the presence of glass. In Fig. 5A—5D, we compare the luminescence emission in two samples, with one using borosilicate glass coverslips, and the other using transparent vinyl coverslips. With all other conditions were maintained the same, we observed strong luminescence signal in the glass sample while none in the vinyl sample. The closeness in distance of the gold nanoparticles is also important. To demonstrate this point, we let IgG-conjugated gold nanoparticles attach to the membrane of the human embryonic cells and took images right on the glass surface and then a little away from the surface. In Fig. 5E—5F, we show that both light-scattering and the nonlinear luminescence were present right on the glass surface; whereas in Fig. 4G—4F, where the focus was adjust 1 μm away from the surface, the luminescence signal diminished to almost zero, while light-scattering was unaffected.

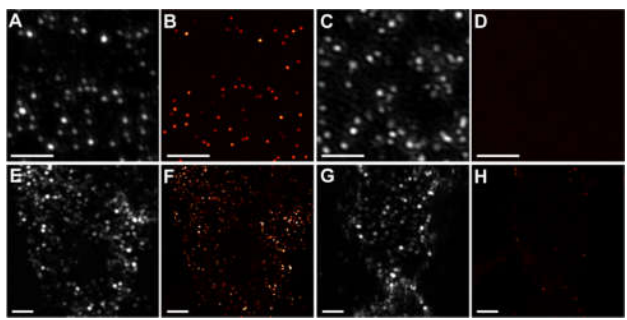

Fig. 5 Nonlinear luminescence was only observed near glass surface. (A and B) Light-scattering (gray-scale) and nonlinear luminescence (red-hot) of gold nanoparticles on a glass surface, are compared to (C and D), where images were taken under exactly the same conditions, but glass was replaced by transparent vinyl, and nonlinear luminescence was not detected. (E and F) Light scattering and nonlinear luminescence from gold nanoparticles attached on a human embryonic kidney (HEK) cell membrane right on a glass surface. (G and H) When focus was changed by 1 μm for the HEK cell, light-scattering was unaffected, whereas nonlinear fluorescence became very weak. . Scale bars represent 1 μm.

The nonlinear luminescence accompanies a strong, nonlinear absorption of the light-scattering. Such absorption is very easy to be observed using the 561 nm laser as excitation (it could be related to that 561 nm is close to the plasmon resonance peak of 70 nm gold nanoparticles). In Fig. 6, each subfigure (A—E) contains a pair of images taken with the same excitation power, in which the upper one shows light-scattering and the bottom one shows luminescence. Fig. 6A—6C show that, when the 561 nm excitation power increased from 0.74 mW to 1.28 mW, a small darkened center appeared in the light-scattering images; meanwhile luminescence was not yet present. When the laser power further increased to 1.76 mW, the darken area grew larger and started to emit at a low level again, and luminescence started to occur. With the 405 nm excitation (Fig. 6D—6E) similar phenomenon were observed. In conclusion, with increasing laser power, it is during or after a darkened center appeared in the light-scattering image when the luminescence starts its emission. This observable photo-darkening suggests that there exists a long-lived intermediate state for the strong nonlinearity more probable to occur.

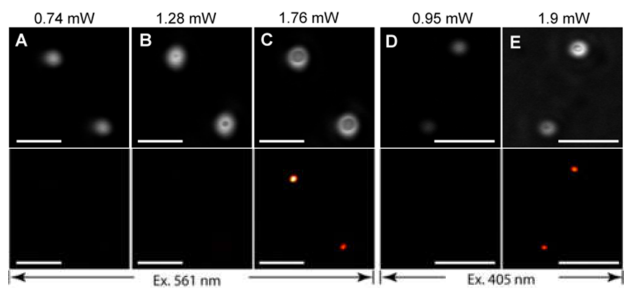

Fig. 6 Nonlinear luminescence follows strong absorption of light-scattering. Scale bars represent 1 μm.

**Discussion**

The nonlinear luminescence we have observed has potential applications in microscopy. On the violet end of the excitation spectrum, it can be used as a simple way to achieve super-resolution. Currently we have demonstrated that the resolution enhancement is 2—3 fold. It is feasible to further boost resolution by exploiting the strong photoblinking [13]. To the red end of the excitation spectrum, it can be very useful in deep tissue imaging if near-infrared excitation is applicable.

**Acknowledgement**

We thank Dr. Rong Lu for preparing biological samples. This contribution was supported by NIH RO1 HL088640 (ES) and NIH RO1 HL107418 (ES & LT).